 \documentclass[manuscript]{aastex}
\usepackage{bm}
\usepackage{amsmath,amssymb}
\usepackage{textcomp}
\usepackage{booktabs}
\usepackage{lscape}
\makeatletter
\newcommand{\tblcaption}[1]{\def\@captype{table}\caption{#1}}
\makeatother
\usepackage{subfigure}

\newcommand\msun {M_\odot}

\begin{document}


\title{OGLE-2008-BLG-355Lb: A Massive Planet around A Late type Star}


\author{N. Koshimoto\altaffilmark{1,A}, A. Udalski\altaffilmark{2,B}, T.Sumi\altaffilmark{1,A}, D.P. Bennett\altaffilmark{3,A}, I.A. Bond\altaffilmark{4,A}, N. Rattenbury\altaffilmark{5}}

\and

\author{F. Abe\altaffilmark{6}, C.S. Botzler\altaffilmark{5}, M. Freeman\altaffilmark{5}, M. Fukagawa\altaffilmark{1}, A. Fukui\altaffilmark{7}, K. Furusawa\altaffilmark{6}, Y. Itow\altaffilmark{6},C.H. Ling\altaffilmark{4}, K. Masuda\altaffilmark{6}, Y. Matsubara\altaffilmark{6}, Y. Muraki\altaffilmark{8}, K. Ohnishi\altaffilmark{9},  To. Saito\altaffilmark{10}, H. Shibai\altaffilmark{1}, D.J. Sullivan\altaffilmark{11}, K. Suzuki\altaffilmark{6}, D. Suzuki\altaffilmark{1}, W.L. Sweatman\altaffilmark{4}, S. Takino\altaffilmark{6}, P.J. Tristram\altaffilmark{12}, K. Wada\altaffilmark{1}, P.C.M. Yock\altaffilmark{5}\\
(MOA Collaboration)}
\author{M.K.Szyma$\acute{\rm{n}}$ski\altaffilmark{2}, M. Kubiak\altaffilmark{2}, I Soszy$\acute{\rm{n}}$ski\altaffilmark{2}, G. Pietrzynski\altaffilmark{2,13}, R. Poleski\altaffilmark{2,14}, K. Ulaczyk\altaffilmark{2}, $\L$. Wyrzykowski\altaffilmark{2,15}\\
(OGLE Collaboration)}

\altaffiltext{1}{Department of Earth and Space Science, Graduate School of Science, Osaka University, 1-1 Machikaneyama, Toyonaka, Osaka 560-0043, Japan}
\altaffiltext{2}{Warsaw University Observatory, Al. Ujazdowskie 4, 00-478 Warszawa, Poland}
\altaffiltext{3}{Department of Physics, University of Notre Dame, Notre Dame, IN 46556, USA}
\altaffiltext{4}{Institute of Information and Mathematical Sciences, Massey University, Private Bag 102-904, North Shore Mail Centre, Auckland, New Zealand}
\altaffiltext{5}{Department of Physics, University of Auckland, Private Bag 92019, Auckland, New Zealand}
\altaffiltext{6}{Solar-Terrestrial Environment Laboratory, Nagoya University, Nagoya, 464-8601, Japan}
\altaffiltext{7}{Okayama Astrophysical Observatory, National Astronomical Observatory, 3037-5 Honjo, Kamogata, Asakuchi, Okayama 719-0232, Japan}
\altaffiltext{8}{Department of Physics, Konan University, Nishiokamoto 8-9-1, Kobe 658-8501, Japan}
\altaffiltext{9}{Nagano National College of Technology, Nagano 381-8550, Japan}
\altaffiltext{10}{Tokyo Metropolitan College of Industrial Technology, Tokyo 116-8523, Japan}
\altaffiltext{11}{School of Chemical and Physical Sciences, Victoria University, Wellington, New Zealand}
\altaffiltext{12}{Mt. John Observatory, P.O. Box 56, Lake Tekapo 8770, New Zealand}
\altaffiltext{13}{Universidad de Concepci\'on, Departamento de Astronomia, Casilla 160-C, Concepci\'on, Chile}
\altaffiltext{14}{Department of Astronomy, The Ohio State University, 140 West 18th Avenue, Columbus, OH 43210,USA}
\altaffiltext{15}{Institute of Astronomy, University of Cambridge, Madingley Road, Cambridge CB3 0HA, United King- dom}
\altaffiltext{A}{Microlensing Observations in Astrophysics (MOA) Collaboration}
\altaffiltext{B}{Optical Gravitational Lensing Experiment (OGLE) Collaboration}


\begin{abstract}
We report the discovery of a massive planet OGLE-2008-BLG-355Lb. The
light curve analysis indicates a planet:host mass ratio of $q = 0.0118 \pm 0.0006$
at a separation of $0.877 \pm 0.010$ Einstein radii. We do not measure a significant
microlensing parallax signal and do not have high angular resolution images that
could detect the planetary host star. Therefore, we do not have a direct measurement
of the host star mass. A Bayesian analysis, assuming that all host stars have
equal probability to host a planet with the measured mass ratio implies a host
star mass of $M_{\rm{h}} = 0.37_{-0.17}^{+0.30} M_{\odot}$  and a companion of 
mass $M_{\rm{P}} = 4.6^{+3.7}_{-2.2}~M_{\rm{J}}$, at a projected separation 
of $r_{\perp} = 1.70^{+0.29}_{-0.30}$ AU. The implied distance to the
planetary system is $D_{\rm{L}} = 6.8 \pm1.1$ kpc. 
A planetary system with the properties
preferred by the Bayesian analysis would be a challenge to the core-accretion model
of planet formation, as the core-accretion model predicts that massive planets
are far more likely to form around more massive host stars. This core accretion
model prediction is not consistent with our Bayesian prior of an equal probability
of host stars of all masses to host a planet with the measured mass ratio. So, if
the core accretion model prediction is right, we should expect that follow-up
high angular resolution observations will detect a host star with a mass in the
upper part of the range allowed by the Bayesian analysis. That is, the host would
probably be a K or G dwarf.
\end{abstract}


\section{Introduction}
Until the first detection of an exoplanet in 1995, planet formation theories referred 
to the formation of the Solar System. The standard core accretion model 
\citep{saf72,hay85,lis93} was believed to be fairly well established, although some 
problems such as the formation of planetesimals \citep[e.g.]{wei93,dom97} remained. 
According to this theory, gas giants, such as Jupiter or Saturn, are formed slightly 
outside the "snow line" where the protoplanetary disk becomes cold enough for water to condense.
However, this theory did not predict the discovery of ``Hot Jupiters"
\citep{may95}, which are planets of about a Jupiter mass with 
orbits lie far inside that of Mercury.
Since then, over 1000 exoplanets (and over 3500 candidates) have been detected. 
The core accretion model now includes the possibility of migration \citep{lin96} to explain
Hot Jupiters, but it still has difficulty
which is derived from the standard model for the origin 
of the solar system is a most generally accepted planet formation scenario, but it 
can't explain all the forms of the exoplanets.
For example, the theoretical prediction of a paucity of the planets with masses of 10 - 100 $M_{\oplus}$ in short period orbits \citep{ida04} is inconsistent with the results from radial velocity studies \citep{how10}. Moreover, today's core accretion model predicts few gas giants orbiting
red dwarf at any separation \citep{lau04,ken08}, and this is confirmed by observations
from radial velocity for massive gas giants orbiting inside the snow line \citep{end06,joh07,cum08,joh10}. 
Early statistical results from the gravitational microlensing method \citep{gou10b,sum10,cas12}
indicate that low-mass, Saturn-like, gas giants are more common around low-mass stars
than Jupiters beyond the snow line, and this is confirmed by radial velocity observations
\citep{montet14}. But, the gravitational microlensing method \citep{mao91,gou92}, has also
revealed several super-Jupiter mass 
planets orbiting just outside of the snow line of their late type host stars \citep{ben06,gau08,don09a,don09b,bat11,kai13,tsa13,shv13} although a quantitative 
analysis of planetary frequency as a function of host star mass has not yet been completed. 
The gravitational microlensing method is capable of discoveries of planets with mass down to the Earth mass just outside of the "snow-line" \citep{ben96}. In terms of the sensitivity region, it is very important for planetary formation theory that the microlensing method is complementary with the other methods, the radial velocity method \citep{but06,bon11} and the transit method \citep{bor11}, which are sensitive to close and relatively massive planets, and the direct imaging method \citep{mar08} which has sensitivity to giant planets with orbital semi-major axes greater than several dozen AU. Also, because microlensing does not rely upon any light from the host star or planet of the lens system for detection \citep{gau12}, it is possible to detect a planet around a star which is too faint to detect by the other methods \citep{ben08} or even a planetary mass object which belongs to no host star \citep{sum11}.

The microlensing events that are searched for planetary signals are discovered by
two microlensing survey groups, the Microlensing Observations in Astrophysics 
group (MOA; \citet{bon01}, \citet{sum03}) and the Optical Gravitational Lensing 
Experiment group (OGLE; \citet{uda03}).
The MOA group uses the very wide field-of-view (2.2 square degrees) MOA-cam3 \citep{sak08}
CCD camera  mounted on the MOA-II 1.8m telescope at the Mt.\ John University 
Observatory in New Zealand. With this large FOV camera, MOA is able to observe 
50 square degrees of our Galactic Bulge every hour, allowing high cadence observations.
MOA detects about 600 new microlensing events and issues alerts of these 
events in real-time every year. 
The OGLE survey is conducted at the Las Campanas Observatory, Chile with 
the 1.3 m Warsaw telescope. In 2008, the OGLE was operating the OGLE-III
survey using the 0.35 square degree OGLE-III camera, but OGLE has now
upgraded to the 1.4 square degree OGLE-IV camera, which enables a higher
cadence survey.

This paper is a report of our analysis of a microlensing event OGLE-2008-BLG-355.
The observations of this event are described in Section \ref{sec-obs}. 
Section \ref{sec-reduction} explains our data reduction procedure.
Section \ref{sec-model} discusses our best model and the comparison with other models.
The source color and the derived the source radius and the Einstein angular radius from the color are 
derivedf in Section \ref{sec-color}.
The likelihood analysis is discussed in Section \ref{sec-like}.
Finally, Section \ref{sec-disc} discusses the results of this work.

\section{Observations}\label{sec-obs}

\begin{figure}
\centering
\epsscale{1.0}
\plotone{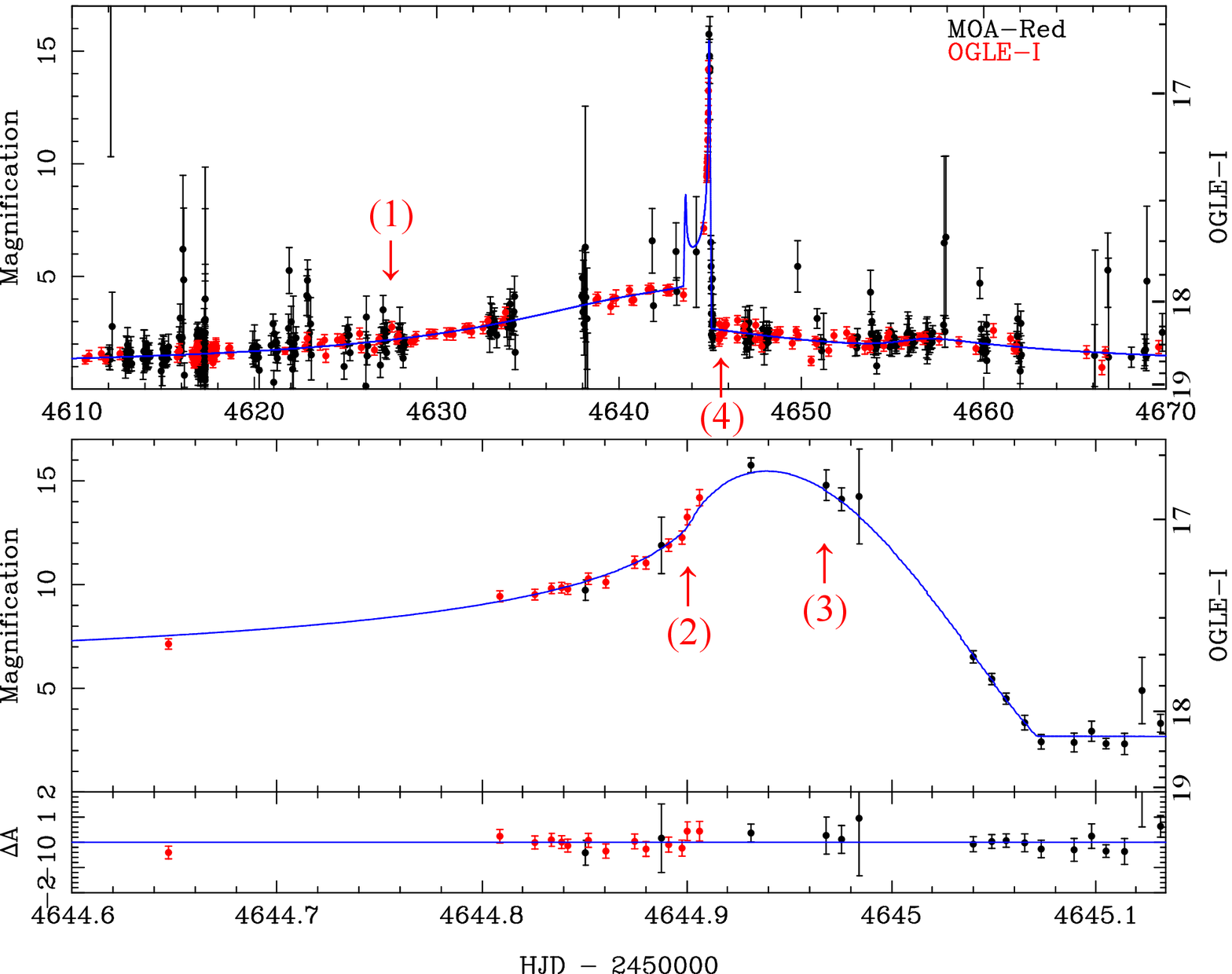}
\caption{The light curve of OGLE-2008-BLG-355 with the best-fit model. The top panel shows the whole event, and the middle panel highlights the caustic exit. The blue line indicates our best-fit standard (i.e. non-parallax) model corresponding to the planetary model parameters in Table \ref{tab-models}. The residuals from the model are shown in the bottom panel. This event was alerted as a microlensing event by OGLE at (1) and its anomaly was alerted at (2). MOA observers increased the cadence of observation of this field from (3) and the data points until then were taken every one hour which is the standard cadence of this field in 2008.}
\label{fig-best}
\end{figure}

Microlensing event OGLE-2008-BLG-355 was detected by the OGLE and MOA
microlensing survey groups.
at $(R.A., Dec.)_{J2000}$ = (17:59:08.81, -30:45:34.1). 
The data are shown in Figure \ref{fig-best}.
The OGLE Early Warning System (EWS) \citep{uda94} alerted this event as 
OGLE-2008-BLG-355 at 2008 UT 19:32 June 9 (HJD$' \equiv $ HJD - 2450000 = 4627.31),
 then in the early morning on June 27, the rising part of the caustic exit was observed 
 by OGLE, and at UT 9:32 on the day (HJD$'$ = 4644.90), OGLE announced
this event as an anomaly event. At UT 3:00 June 28 (HJD$'$ = 4645.63), the 
following day the OGLE anomaly alert, MOA also independently found this event and alerted 
the event as MOA-2008-BLG-288. The OGLE observations were made primarily in the $I$-band
while the MOA observations were made in the custom MOA-Red filter which is similar to
the sum of the standard Cousins $R$ and $I$-band filters.

The OGLE anomaly alert (HJD$'$ = 4644.90) was announced about an hour prior to the 
center of the peak of magnification (HJD$' \simeq$ 4644.94). As a result, the MOA observers 
increased the cadence of observation of this field from HJD$' \simeq$ 4644.97 
which provided good sampling around the caustic exit. MOA observed the event with 
its standard one-hour cadence immediately prior to the second caustic crossing peak. 
Following the OGLE anomaly alert, MOA used a higher cadence during the time after
the second caustic peak and during the caustic exit. Because of this strategy, MOA was
able to measure the source angular radius parameter, $\rho$. The best fit 
microlensing model parameters for this event are shown in Table \ref{tab-models}. 
The alert history is also shown in the light curve Figure, (Figure~\ref{fig-best}) 
which also shows our best-fit microlensing model, discussed in Section \ref{subsec-comp}.
The caustic entry was not observed by either OGLE or MOA.

\section{Data Reduction} \label{sec-reduction}
\begin{figure}
\centering
\epsscale{0.7}
\plotone{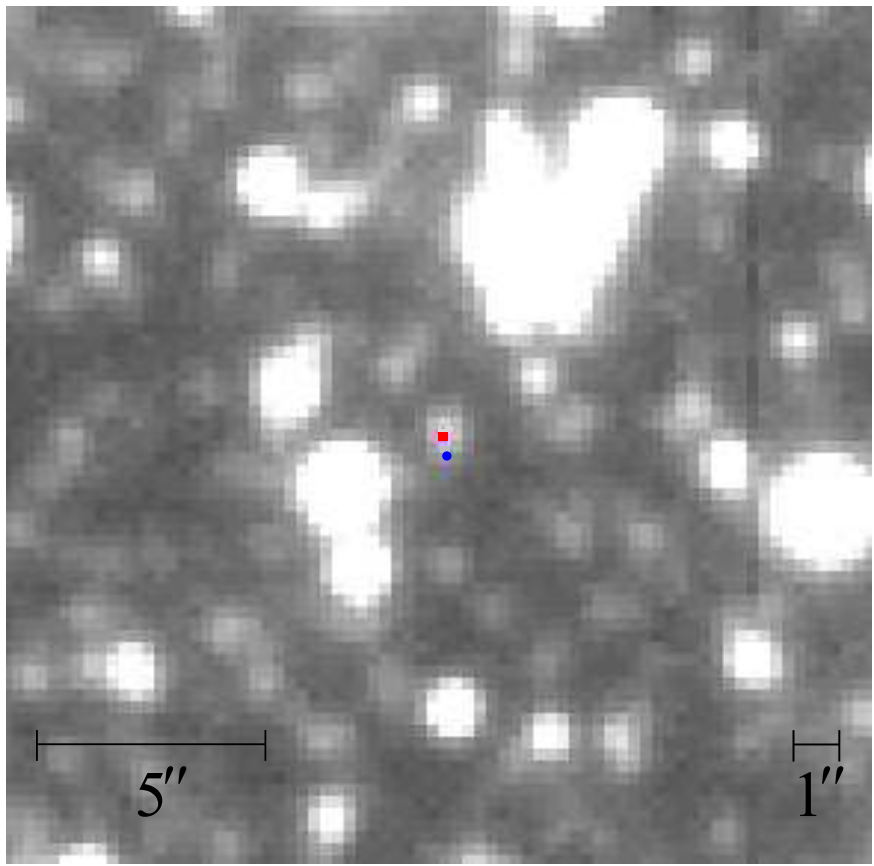}
\caption{OGLE $I$-band image around the source star. The blue circle and the red square dots indicate the source star (not magnified) and an unresolved brighter star respectively. The difference between them is as large as 0.23$''$. We can also find several bright stars near the source.}
\label{fig-nearby}
\end{figure}

There are bright stars near the source star of this event. Figure \ref{fig-nearby} shows them on the OGLE $I$-band image. Therefore, both OGLE and MOA photometry data of the target star are affected by these nearby bright stars. The influence on the OGLE data appears as a centroid shift of the target. OGLE data are reduced by the OGLE Difference Image Analysis (DIA) photometry pipeline \citep{uda03}. In the OGLE online data, the center of the resolved star on the reference image is used as the centroid for a PSF fit. In this event, the center of faint source star is slightly shifted to the cataloged bright star. Thus we re-reduced the OGLE data with the correct centroid.  
MOA data were reduced by the MOA DIA pipeline \citep{bon01} and given in the form of $\Delta$Flux which is defined as the residual flux from the reference flux for the DIA.
From these, we obtain 1425 OGLE $I$-band (hereafter OGLE $I$) data and 7735 MOA-Red band data.
In this section, we reduce these obtained data further as described below. The influence of the nearby stars on the MOA data is described in Section \ref{sec-seeing}.

\subsection{Systematics} \label{sec-seeing}
\begin{figure}
\epsscale{.70}
\rotatebox{-90}{
\plotone{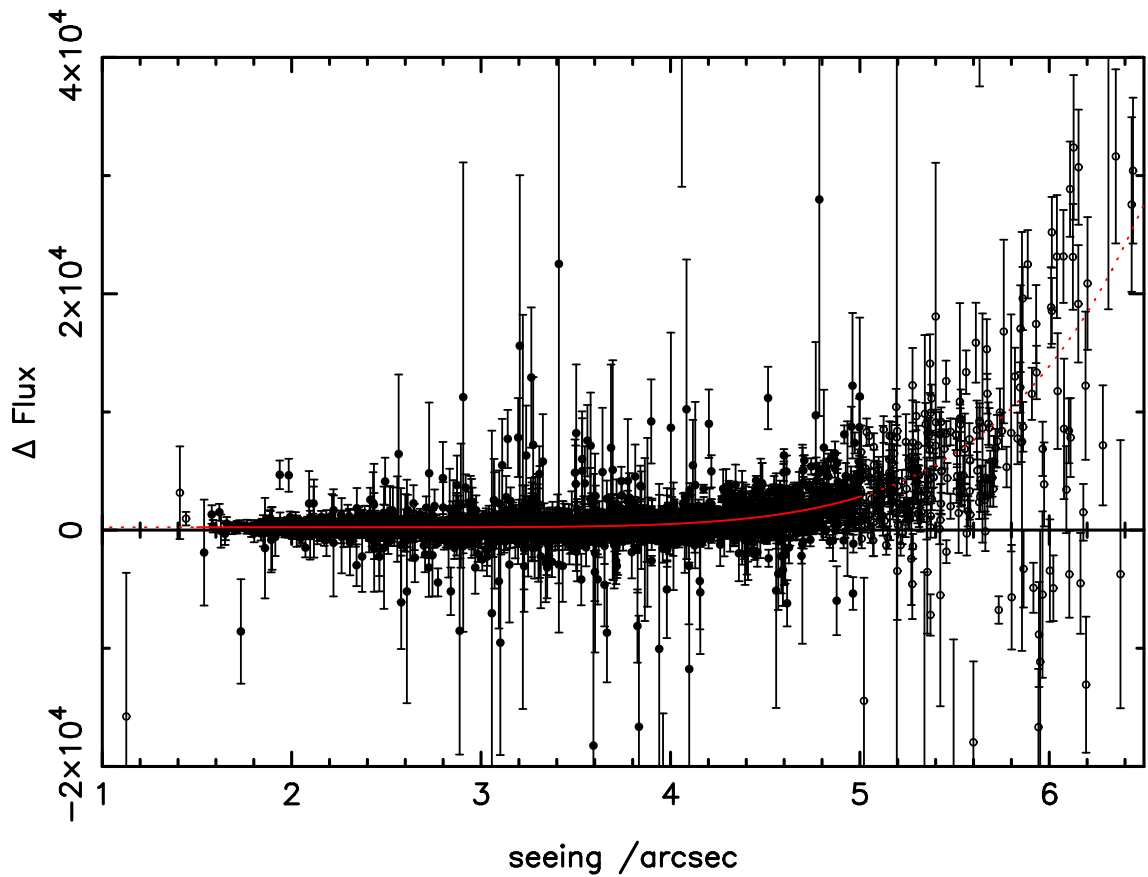}
}
\caption{The relationship between seeing and $\Delta$Flux in MOA data. The filled and open circles indicate data within and outside of the range 1.5 arcsec $\leq$ seeing $\leq$ 5 arcsec, respectively. Only filled circles within this range are used to derive the best fit curve (red solid line) given by Equation(\ref{eq-seeing}). The solid and dotted red lines indicate the best fit model inside and outside of the fitted range 1.5 arcsec $\leq$ seeing $\leq$ 5 arcsec, respectively.}
\label{fig-seeing}
\end{figure}

MOA photometry for this event includes extra flux under an influence of nearby bright stars depending on seeing. Figure \ref{fig-seeing} shows the $\Delta$Flux of the event's baseline as a function of seeing. We can see a tendency that as seeing increases, the larger delta flux value becomes.
We find the best fit empirical relation in Figure \ref{fig-seeing} to be:
\begin{align}
\Delta \rm{Flux} = 238.43 + 0.65({\rm seeing} - 1.5)^{6.62}. \label{eq-seeing}
\end{align}
for data with ${\rm seeing} \geq 1.5$ arcsec. 
We apply the seeing corrections to the MOA $i$th photometry data point by using
\begin{align}
\Delta {\rm Flux} _i'= \Delta {\rm Flux}_i - 0.65({\rm seeing}_i - 1.5)^{6.62}, \label{eq-seeing-corre}
\end{align}
where we removed 363 data points with seeing outside the range 1.5 arcsec $\leq$ seeing $\leq$ 5 arcsec. 
The OGLE data are not affected by extra flux because the seeing values for the OGLE-III data are smaller than these for the MOA-II data.

As will be mentioned in Section \ref{subsec-para}, there are other systematic errors in the baseline of both the OGLE $I$ and MOA-Red data which imitate the perturbations caused by the parallax effect.
Therefore we use data from 2008 in our analysis. 
There is enough baseline data in 2008 for this event because the event is not too long and occurred in the middle of the 2008 bulge season. 
Our final data set comprises 336 OGLE $I$ data points and 1112 MOA-Red data points.

\subsection{Error Normalization}
It is generally known that the photometry errors given by photometry codes are underestimated \citep{yee12}. The error bars for the data points have been re-normalized such that the reduced $\chi^2$ of the best-fit model $\chi^2 / dof \simeq 1$.
For re-normalizing, we used the standard formula
\begin{align}
\sigma _i' = k \sqrt{\sigma^2_i + e_{min}^2} \label{eq-reno}
\end{align}
where $\sigma_i$ is the original error of the $i$th data point in magnitudes, and the re-normalizing parameters are $k$ and $e_{min}$.
This nonlinear formula operates so that the error bars at high magnification, which can be affected by flat-fielding errors, can be corrected by $e_{min}$.
These parameters $e_{min}$ are adjusted so that the cumulative $\chi^2$ distribution as a function of the number of data points sorted by each magnification of the best model is a straight line of slope 1.
We found $e_{min} = 0,~ k = 1.403$ in MOA-Red and $e_{min} = 0.01149, k = 1.213$ in OGLE $I$ and thereby corrected the errors using formula (\ref{eq-reno}).






\section{Modeling} \label{sec-model}
In the microlensing method, the parameters of the lens object can be obtained by fitting a microlensing model to the data.
The fitting parameters for a standard binary lens model are the Einstein radius crossing time, $t_E=\theta_E/\mu_{rel}$, where $\theta_E$ and $\mu_{rel}$ are the angular Einstein radius and the lens-source relative proper motion respectively, the time, $t_0$, when the source is closest to a reference point, the source's closest approach, $u_0$, to the reference point on the lens plane at time $t_0$ in units of the Einstein radius, the secondary-primary mass ratio, $q$, the projected separation between lens objects in Einstein radius units, $s$, the angle of the source trajectory with respect to the binary lens axis, $\alpha$, and the angular radius of the source star ($\theta_*$) relative to the angular Einstein radius ($\theta_E \equiv R_E/D_L$), $\rho \equiv \theta_*/\theta_E$.
With the magnification variation against time, $A(t,\bm{x})$, which is defined in terms of the above parameters $\bm{x}=(t_E,t_0,u_0,q,s,\alpha,\rho)$, we can linearly fit
\begin{align}
F(t) = f_s A(t,\bm{x}) + f_b \label{eq-F}
\end{align}
to a data set and obtain the instrumental source flux $f_s$ and the instrumental blending flux $f_b$ for every telescope and pass-band.

We search the best-fit parameters using this standard binary model and compare the best model with other standard models in Section \ref{subsec-comp}. Then, in Section \ref{subsec-para}, we discuss the significance of the parallax effect which is one of higher order microlensing effects and show that a standard binary model is preferred over a parallax model for this event.

\subsection{Limb Darkening}
When a point source object passes a caustic line, the magnification of the source diverges to infinity. But because the source object has a finite extent, a light curve has a finite peak even if the source passes a caustic. Conversely, we can obtain the finite source star parameter, $\rho \equiv \theta_*/\theta_E$, by analyzing the peak of a caustic crossing and this allows us to break one of the degeneracies between the lens properties.

In this event, the caustic exit was observed at high cadence by OGLE and MOA. When finite source effects are important in this event, the limb darkening effect must be included in the modeling to obtain the proper model.
We adopt a linear limb-darkening law with one parameter for the source brightness:
\begin{align}
S_{\lambda} (\vartheta) = S_{\lambda}(0)[1-u_{\lambda}(1-\cos \vartheta )]. \label{eq-limb}
\end{align}
Here, $\vartheta$ is the angle between the normal to the stellar surface and the line of sight, $S_{\lambda} (\vartheta)$ is the brightness from the source at the orientation of $\vartheta$ and $u_{\lambda}$ is the limb-darkening coefficient.
According to \citet{gon09}, we estimate the effective temperature, $T_{\rm{eff}}\sim 5803$ K from the source color which is discussed in Section \ref{sec-color} and assumed a metallicity of log$~[M/H]=0.0$. With $T_{eff} = 5750 K$ and assuming surface gravity $\log~ g = 4.5~ \rm{cm ~s^{-2}}$ and log$~[M/H] = 0.0$, the limb-darkening coefficients selected from \citet{cla00} are $u_I=0.5290,~ u_R=0.6114$. Therefore we used the $u_I$ for OGLE $I$ and the mean of the $u_I$ and $u_R$, 0.5702 for MOA-Red, the filter which has the range of both the standard $I$ and $R$ filters.

\subsection{The Best-Fit Model}\label{subsec-comp}
This event has been already published as a brown dwarf event with mass ratio of $q = 0.106$ in \citet{jar10} (hereafter JA10). They analyzed systematically OGLE archival binary events using only OGLE data. In 2012, MOA also conducted a systematic search for MOA archival data as well and found a preference for a planetary model when including MOA data in this event. This is the context of this work and in this section, we confirm the best planetary model and compare the model with other models which have mass ratios $q$ in the range of $-4 < {\rm log}~q < \rm{0}$. Note that the OGLE data that JA10 used for this event is different from the one that we used because we re-reduced the data for this analysis as mentioned in section \ref{sec-reduction}.

\begin{figure}
\epsscale{0.8}
\plotone{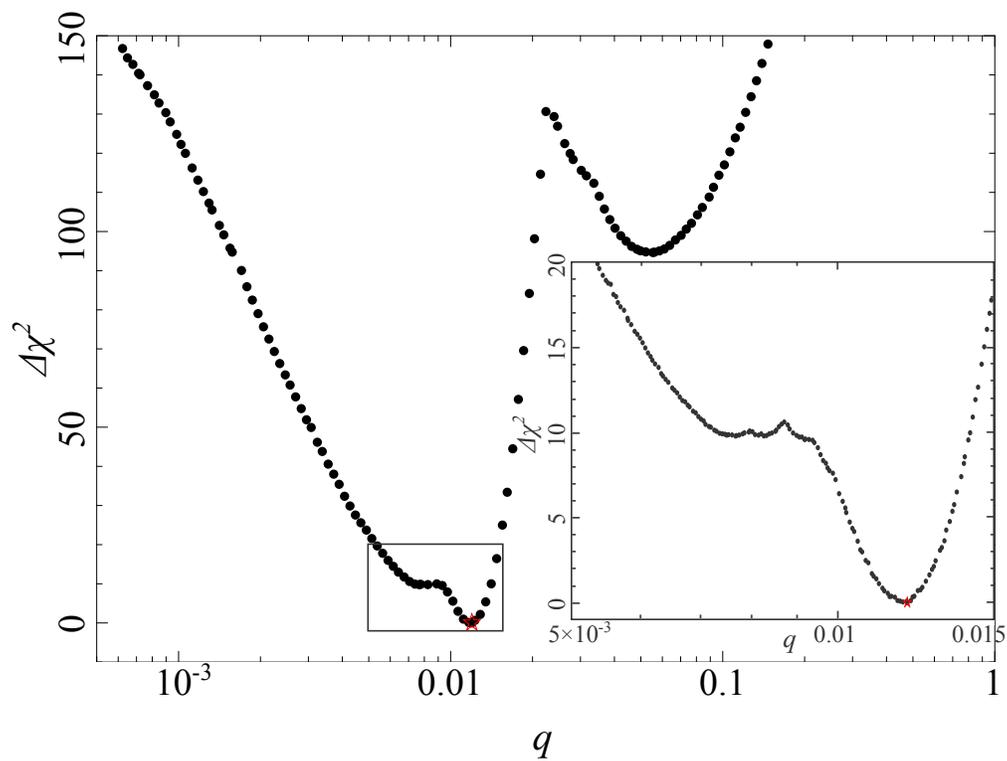}
\caption{$\Delta\chi^2$ as a function of mass ratio $q$. $\Delta \chi^2$ means the difference from the minimum $\chi^2$ at $q = 1.18 \times 10^{-2}$ (the red star model). The inner figure is a close-up of the part around the best model enclosed by the gray square. The size of a bin is $\Delta (\rm{log}~ \it{q}) = \rm{0.02}$ in the outer figure and it is $0.003$ in the inner figure.}
\label{fig-q-chi2}
\end{figure}

In order to find the model which has the smallest $\chi^2$ value, we used a Markov Chain Monte Carlo (MCMC) approach \citep{ver03}, the image centered ray-shooting method \citep{ben96,ben10}, starting from a large number of initial values of gridded over the wide parameter space at a number of fixed $q$ values. Figure \ref{fig-q-chi2} shows the $q$ vs $\Delta\chi^2$ plot with respect to the models which have $\Delta \chi^2$ less than 150. Here, $\Delta \chi^2$ means the difference of $\chi^2$ between each model and the best model.
From Figure \ref{fig-q-chi2}, we find that the best model locates around a planetary mass ratio $q \sim 1.2 \times 10^{-2}$ and there are broadly 2 other local minima in this $q$ range, around $q \sim 8 \times 10^{-3}$ with $\Delta \chi^2 \sim 10$ in around which a few small dips exist and $q \sim 5.5 \times 10^{-2}$ with $\Delta \chi^2 \sim 95$.
The best model in the range of $q > 0.02$, which does not correspond to a planetary mass ratio generally, is a local minimum of $q \sim 5.5 \times 10^{-2}$ corresponding to the model given by JA10. Note this model is slightly different because we used of re-reduced OGLE data and MOA data.

\begin{figure}
\epsscale{0.9}
\plotone{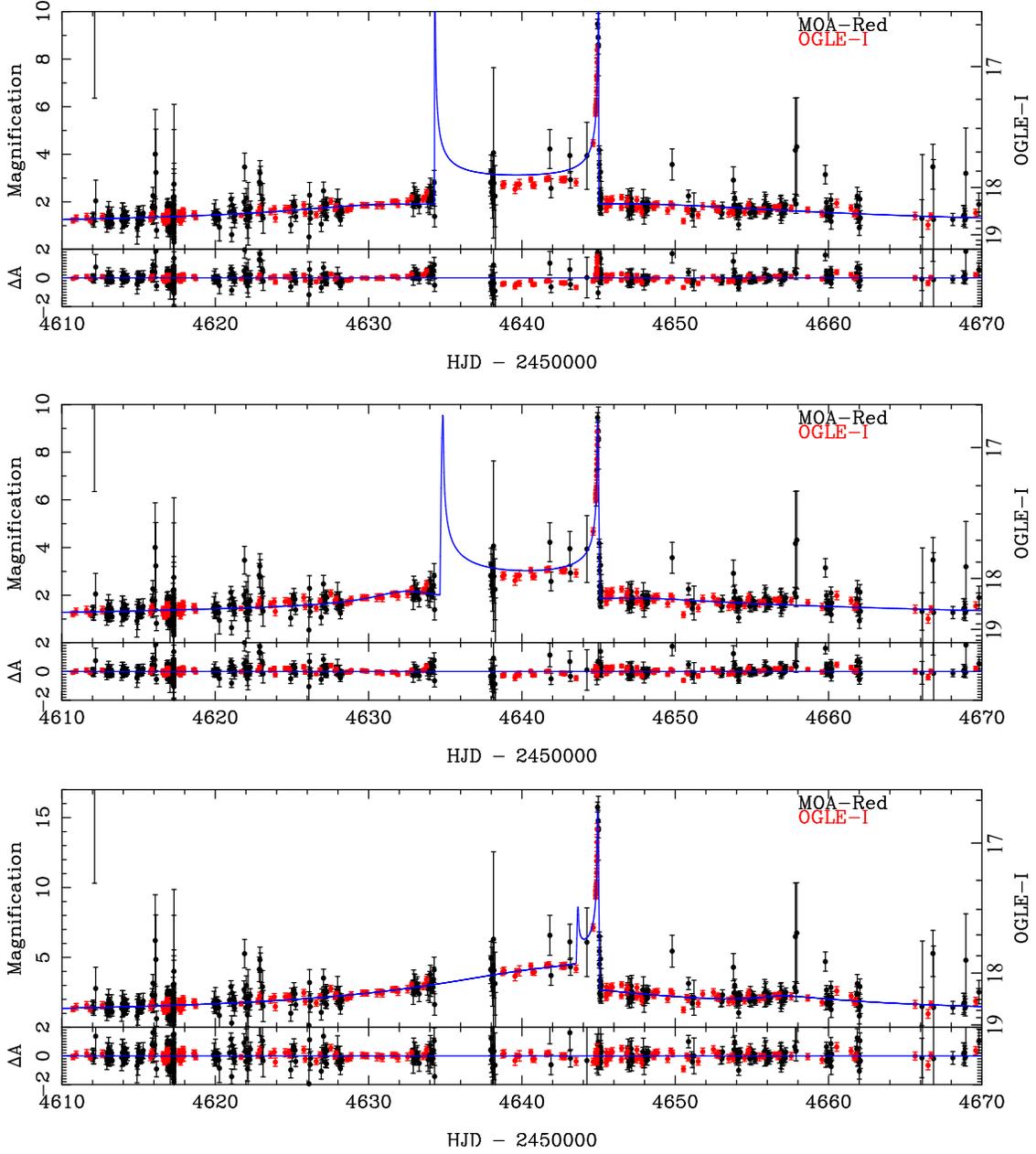}
\caption{The light curves with the models in Table \ref{tab-models}. The top is the best model in JA10, the middle is the best brown dwarf model and the bottom is the best planetary model found in this work. Each curve in blue corresponds to the model parameters given in Table \ref{tab-models}}.
\label{fig-models}
\end{figure}

\begin{figure}
\centering
\epsscale{0.7}
\rotatebox{-90}{
\plotone{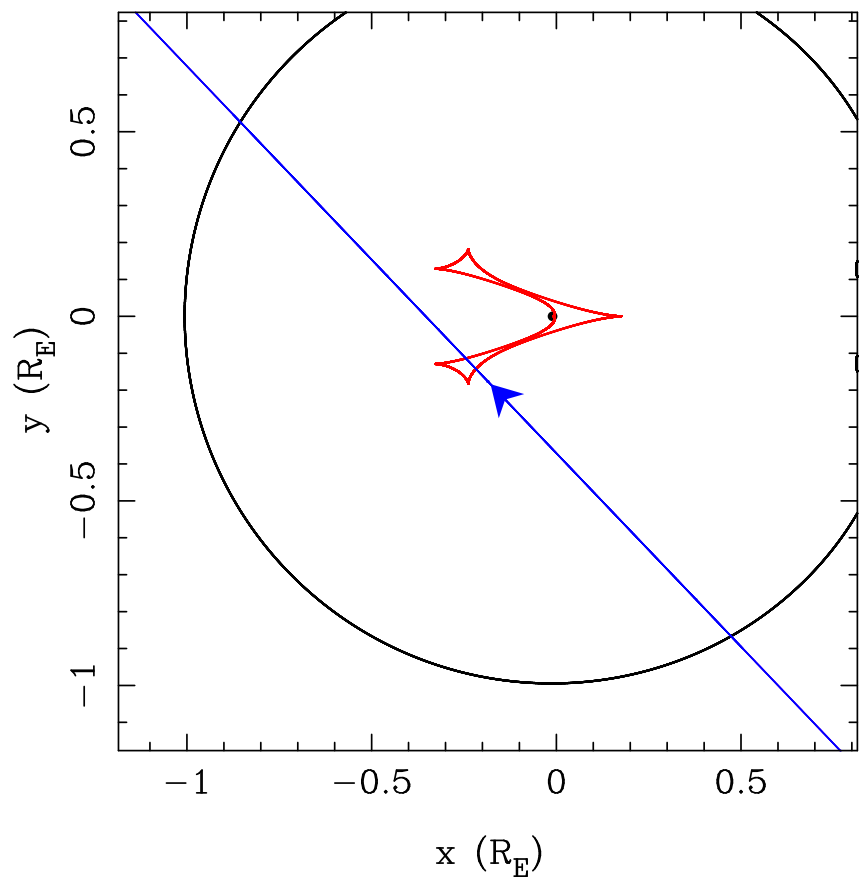}
}
\caption{The caustic curve (red line) plotted for the OGLE-2008-BLG-355 best-fit model. The blue line indicates the source trajectory and the black curve is the critical curve.}
\label{fig-cau}
\end{figure}

The light curve with the model of JA10 is shown in the top panel of Figure \ref{fig-models}, and the middle and bottom panels show the best brown dwarf model and the best planetary model respectively, which are obtained by letting $q$ float as a free parameter. The parameters of these models are shown in Table \ref{tab-models} and the caustic of the best planetary model is shown in Figure \ref{fig-cau}.
To reproduce the brown dwarf model in JA10, we used their result as the initial starting point and the parameters $q, s, t_E$ and $\rho$ were fixed, and $t_0,~ \alpha$ and $u_0$ were used as free parameters because the definitions of these parameters are defined differently. Note that $\rho$ is fixed to zero for this event in JA10.
As to these light curves, the preference is also visible in the shape of the caustic interior at HJD$'$ = 4638 - 4644 and furthermore the mini-bump around HJD$'$ = 4658 found in only the planetary model is confirmed by both OGLE and MOA consistently.

This event prefer the planetary model to the model corresponding to JA10. This is likely because of the added MOA data and the optimized data treatment detailed in Section \ref{sec-reduction}.
The comparison between the best planetary and brown dwarf models was also conducted using only OGLE data and only MOA data in separate analyses, and we obtained the same order of preference in both cases. With OGLE alone, the $\Delta \chi^2$ between the best planetary model and the best model in $q > 0.02$ is about 54 and the $\Delta \chi^2$ value becomes about 36 with MOA data alone.

\begin{figure}
\centering
\epsscale{0.7}
\plotone{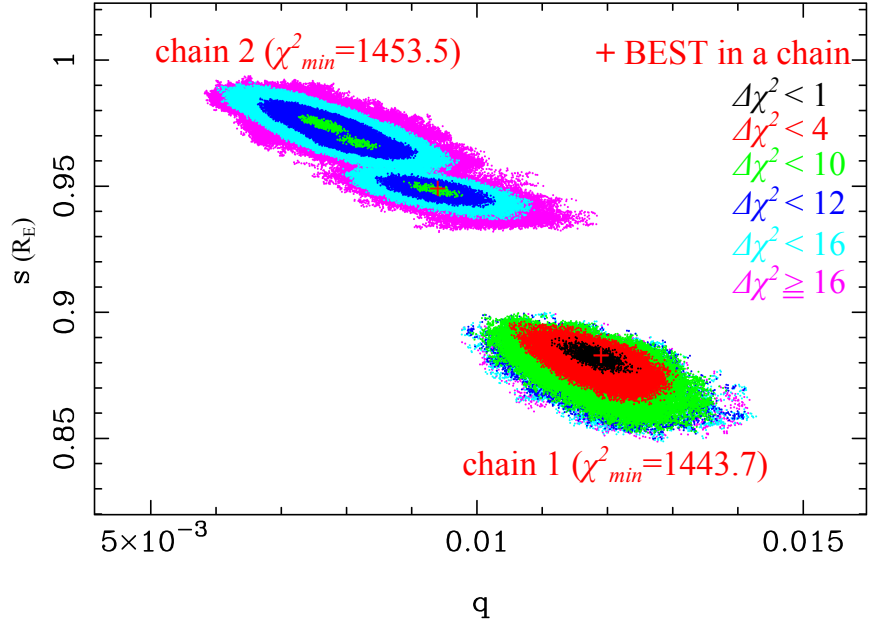}
\caption{The $\chi^2$ distribution of 2 chains of MCMC in $q$ vs $s$. Chain 1 is the chain of MCMC around the best model and chain 2 is that of the local minimum of $\Delta \chi^2 = 9.8$. The points are color coded based on their $\Delta \chi^2$, the difference from the $\chi^2$ minimum of 1443.7, according to the ranges shown in the upper right in the figure. Here 10 or 12 are used as the values of the borders of $\Delta \chi^2$ for clarity, in order to emphasize the 3 local minimums in chain 2, which are consistent with the dips in the plateau in the range of $7.3 \times 10^{-3} < q < 9.3 \times 10^{-3}$ in Figure \ref{fig-q-chi2}. The red crosses locate in the minimum $\chi^2$ in each chain, $\chi^2_{min} = 1443.7$ in chain 1 and $\chi^2_{min} = 1453.5$ in chain 2.}
\label{fig-chain}
\end{figure}

On the other hand, to verify the shape of $q$ vs $\chi^2$ plot in Figure \ref{fig-q-chi2} around $q \sim 8 \times 10^{-3}$ and $\Delta \chi^2 \sim 10$, which looks almost flat but has a few small dips, we check the MCMC chains for the best model and the local minimum. Figure \ref{fig-chain} shows the $\chi^2$ distribution of the chains in $q$ vs $s$. From this figure, we find that there are 4 local minima in the range $6 \times 10^{-3} < q < 1.4 \times 10^{-2}$ and one of this is the best model at $q = 1.18 \times 10^{-2}$. The 3 other minima are located around $q \sim 7.5 \times 10^{-3}$, $8 \times 10^{-3}$ and $9.5 \times 10^{-3}$ and these locations are consistent with the locations of dips in Figure \ref{fig-q-chi2}. Therefore, we can verify the shape around $q \sim 8 \times 10^{-3}$ in Figure \ref{fig-q-chi2} and find that there are 3 other planetary models which have $\Delta \chi^2 \sim 10$.

From the above results, we conclude that this event is best explained by a planetary model which has $\chi^2 = 1443.7$ and a mass ratio $q = (1.18 \pm 0.06) \times 10^{-2}$. The planetary parameter values are shown in Table \ref{tab-models} and we use them in the following discussion. Figure \ref{fig-best} shows the light curve of this event with this best-fit model.

\subsection{Parallax Model} \label{subsec-para}
When the event time scale $t_E$ is relatively long, typically $t_E > 50$ days, the light curve can be affected by the difference between the parallax of the source and that of the lens. Then, we can measure a new physical quantity, $\pi_{E,N},\pi_{E,E}$, which are respectively the north and east component of the relative parallax vector between the source and the lens, $\bm{\pi_E}$ \citep{gou00}. This is known as the microlensing parallax effect. By obtaining the parallax parameter, $\pi_E = \sqrt{\pi_{E,N}^2+\pi_{E,E}^2}$, the finite source effect parameter, $\rho$, and the source angular radius, $\theta_*$, the degeneracies of lens properties in $t_E$ is broken entirely, i.e., one can calculate the mass, $M_L$, the distance, $D_L$, and the relative proper motion, $\mu_{rel}$, of the lens star system.

\begin{figure}
\centering
\epsscale{0.6}
\rotatebox{-90}{
\plotone{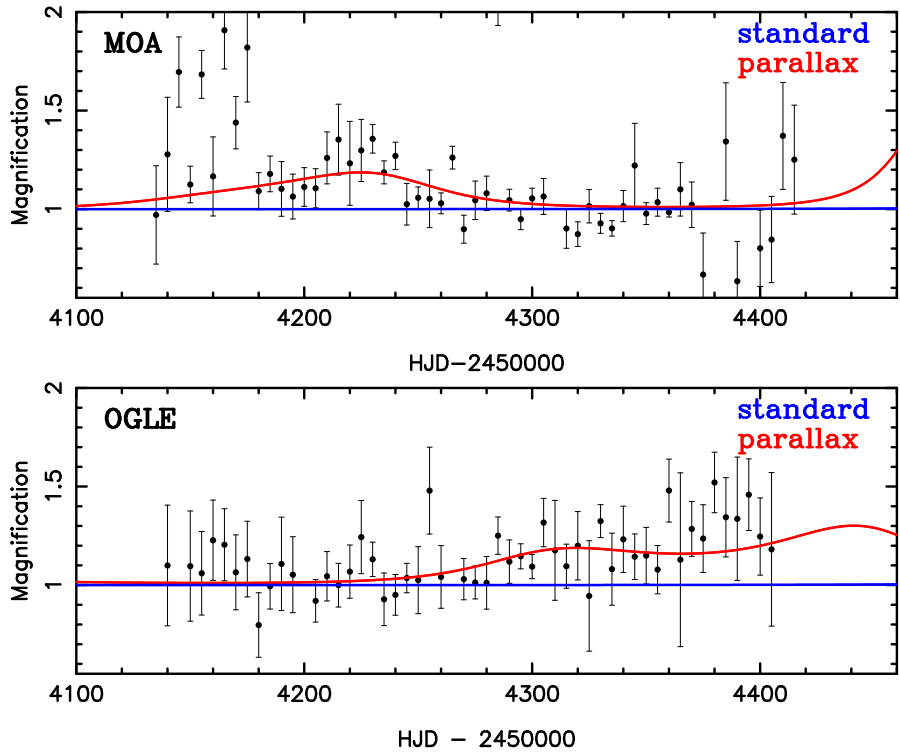}
}
\caption{Zoom on the data points from 2007. Light curves are shown of MOA-Red data (top panel) and OGLE data (bottom panel) with the best standard model (blue line) and parallax models obtained by fitting with only either MOA or OGLE data (red line).
Here both of MOA and OGLE data are binned by 5 days. MOA and OGLE data are inconsistent which indicate the variations are not real.}
\label{fig-bin}
\end{figure}

In this event, using all data available, not only the 2008 data as the previous discussion above, we searched for the best parallax model and found that the $\chi^2$ value is improved by about 70 over the non-parallax model.
However, the most of the parallax signal came from an unexpected part of the light curve, the baseline of the previous year. Therefore, we analyzed the MOA data or OGLE data separately in order to check whether the parallax signal came from the same part of the data sets or not. In the MOA data, the parallax signal came from the first half of previous year, while in the OGLE data, the parallax signal came from the last half of the previous year. Figure \ref{fig-bin} shows the light curves of MOA and OGLE data points from the previous year by binning of 5 days. We found that the MOA and OGLE data were clearly inconsistent in the previous year. Next, we found fits for each data set, removing the data points of previous year. However this also resulted in parallax parameters inconsistent with each other. Therefore, we conclude that the measured parallax signal is not real in this event and analyze this event using only 2008 data to prevent the systematic errors in the base line from making mischief. With only 2008 data, the $\Delta \chi^2$ between the parallax and non-parallax model become 0.36 and thus we could not detect a parallax signal.

\section{The Angular Einstein Radius} \label{sec-color}
To perform the likelihood analysis of Section \ref{sec-like}, we use the event time scale $t_E$ and the angular Einstein radius $\theta_E$ as observed values. In order to yield $\theta_E=\theta_*/\rho$, not only $\rho$, which can be obtained as the one of the fitting parameters, but also the angular source radius $\theta_*$ is required. $\theta_*$ can be estimated from the source color, $(V - I)_S$, and the magnitude, $I_S$, empirically \citep{ker04}.

\subsection{Source Color and Magnitude}
In the case of OGLE-2008-BLG-355, no $V$-band data were taken because the event was not recognized as an event involving a planet signal until our analysis in 2012. Therefore, we estimated the source color by using the other method proposed by \citet{gou10}.
This method yields $(V - I)$ by means of the slight difference in wavelength between MOA-Red and OGLE $I$. We find the approximate linear relation of $(V - I)_{\rm{OGLE}}$ to $(I_{\rm{OGLE}} - R_{\rm{MOA}})$ using isolated field stars around the source star. Then, using the value of $(I_{\rm{OGLE}} - R_{\rm{MOA}})$ of the source star gained from the best-fit model, we can get the $(V - I)_{\rm{OGLE}}$ of the source star.


First, we derive $(I_{\rm{OGLE}}-R_{\rm{MOA}})$ of the source star from the best-fit model. In order to compare with the field stars later, the source magnitude in the same scale with that of the field stars must be obtained.
Then we make the light curve using DoPHOT \citep{dophot} because photometry of the field stars of MOA are done by using DoPHOT in the next step. In dense fields such as those toward the bulge, the accuracy of differential photometry by DIA is better than that of DoPHOT. Hence we make the light curve using DIA with the same PSF as DoPHOT to obtain the source magnitude in the DoPHOT scale but having the accuracy of DIA photometry. 
The instrumental source flux can be obtained by the linear fit of Equation (\ref{eq-F}) for the parameters, $\bm{x}$, of the best-fit model and then, we obtain
\begin{align}
(I_{\rm{OGLE,light}} - R_{\rm{MOA,DoPHOT}})_S = 2.87 \pm 0.02. \label{eq-IRs}  
\end{align}
Here, the index of "light" and "DoPHOT" represent the both scales of instrumental magnitude of OGLE light curve (hereafter OGLE-light scale) and DoPHOT respectively and the "$S$" denotes the source star.

\begin{figure}
\centering
\epsscale{0.7}
\plotone{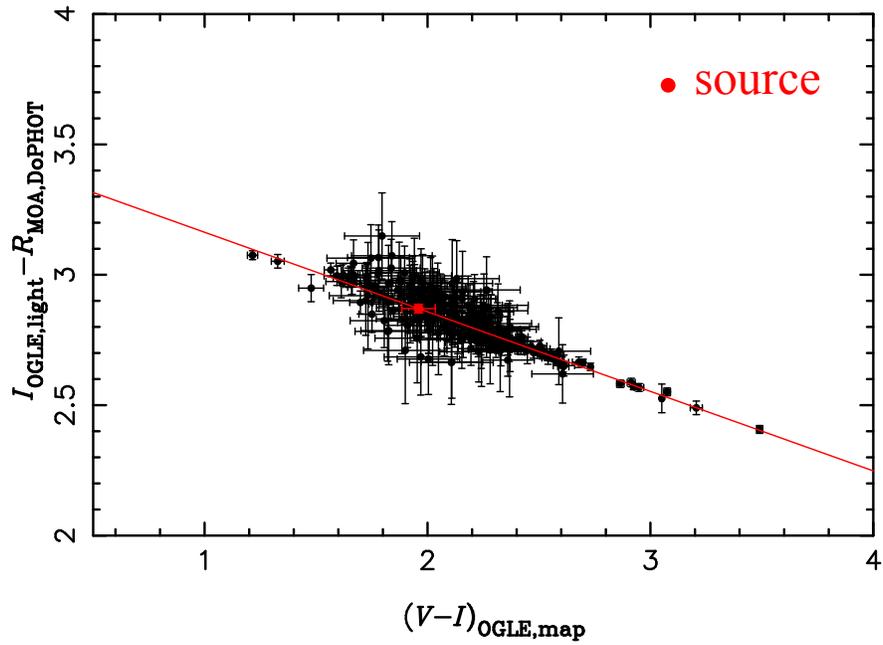}
\caption{The relation between $(I_{\rm{OGLE,light}} - R_{\rm{MOA,DoPHOT}})$ and $(V - I)_{\rm{OGLE,map}}$ in the isolated stars within 2$'$ around the source star. Here $2.5\sigma$ outliers are recursively rejected. The red line and point indicate respectively the best-fit of linear model of Equation (\ref{eq-IR-VI}) and the value of the source star.}
\label{fig-IR-VI}
\end{figure}

Next, we get the relation of $(V - I)_{\rm{OGLE}}$ to $(I_{\rm{OGLE}}-R_{\rm{MOA}})$. $R_{\rm{MOA}}$ values are obtained from MOA reference images by using DoPHOT and $I_{\rm{OGLE}}$ and $V_{\rm{OGLE}}$ are obtained from the OGLE-III photometry map \citep{szy11}.
We plot these values for stars within 2$'$ around the source star in Figure \ref{fig-IR-VI} in which the vertical axis is $(I_{\rm{OGLE,light}}-R_{\rm{MOA,DoPHOT}})$ and the horizontal axis is $(V - I)_{\rm{OGLE,map}}$. The index of "map", against that of "light", is used for the magnitude in OGLE-III photometry map scale (hereafter OGLE-map scale). 
The relations between the magnitude in the scale of OGLE-map, $V_{\rm{OGLE,map}}$, $I_{\rm{OGLE,map}}$, and OGLE-light, $V_{\rm{OGLE,light}}$, $I_{\rm{OGLE,light}}$, are given as
\begin{align}
(V - I)_{\rm{OGLE,map}} &= 0.925(V - I)_{\rm{OGLE,light}}\label{eq-cali1}\\
I_{\rm{OGLE,map}} &= I_{\rm{OGLE,light}} + 0.039(V- I)_{\rm{OGLE,map}} .\label{eq-cali2}
\end{align}
We must add the following additional correction if the calibrated color $(V - I)_{\rm{OGLE,map}}$ is larger than 1.5 mag \citep{szy11}.
\begin{align}
\Delta I_{\rm{OGLE,map}} = -0.033918 + 0.016361(V - I)_{\rm{OGLE,map}} + 0.004167(V - I)_{\rm{OGLE,map}}^2.\label{eq-cali3}
\end{align}
The photometry and values were treated according to the following.

DoPHOT photometry of MOA stars is likely to include "extra flux" relative to the corresponding OGLE-III stars, because the MOA pixel size is about
twice as large as the OGLE pixel size and the seeing of MOA data is about 1.5 times larger than OGLE. Therefore, if there are other stars within $1.8''$ from the target star in the OGLE-III photometry map, we did not include them in Figure \ref{fig-IR-VI}. This way ensures that the calibration is done only using the isolated stars.
With respect to $I_{\rm{OGLE}}$ in the vertical axis in Figure \ref{fig-IR-VI}, we converted the magnitudes in OGLE-map scale to OGLE-light scale by using the relations of Equations (\ref{eq-cali1}) - (\ref{eq-cali3}) because the obtained source magnitude $I_S$ in Equation (\ref{eq-IRs}) was in the OGLE-light scale.
Next, we fitted them to a function of the form, ($I_{\rm{OGLE,light}} - R_{\rm{MOA,DoPHOT}}) = a + b~(V - I)_{\rm{OGLE,map}}$, and recursively removed 2.5 $\sigma$ outliers.
We also removed the handful of stars with $(V - I) > 4$ or $\sigma_{(V - I)} > 0.2$ because they are very far from our range of interest and show slightly larger scattering (although this cut hardly affects the calculation). Thus, we obtained an equation,
\begin{align}
(I_{\rm{OGLE,light}} - R_{\rm{MOA,DoPHOT}}) = (3.47 \pm 0.01) - (0.305 \pm 0.005)(V - I)_{\rm{OGLE,map}}.\label{eq-IR-VI}
\end{align}

Finally, by assigning Equation (\ref{eq-IRs}) to Equation (\ref{eq-IR-VI}), we derive the color of the source star,
\begin{align}
(V - I)_S \equiv (V - I)_{\rm{OGLE,map},\it{S}} = 1.96 \pm 0.07. \label{eq-VIs} 
\end{align}
Moreover, $I_{\rm{OGLE,light},\it{S}}$ and $I_{\rm{OGLE,light},\it{b}}$, which are obtained by Equation (\ref{eq-F}) with the best-fit model, can be calibrated by Equation (\ref{eq-cali1}) - (\ref{eq-cali3}) and we get
\begin{align}
I_S \equiv I_{{\rm OGLE,map},S} = 20.02 \pm 0.12 \label{eq-Is} \\ 
I_b \equiv I_{{\rm OGLE,map},b} = 19.24 \pm 0.07 \label{eq-Ib}  
\end{align}
as the source and blending magnitude in OGLE-map scale. Here, the resolved star magnitudes cataloged in the OGLE-III photometry map \citep{szy11} are 
\begin{align}
(V , I)_{cata} = (21.124, 18.794) \pm (0.268, 0.148). \label{eq-V,Iref}
\end{align}
Hence, we applied 
\begin{align}
(V - I)_{b} &\equiv (V - I)_{{\rm OGLE,map},b}\notag\\
            &= -2.5 ~{\rm log} (10^{-\frac{V_{cata}}{2.5}} - 10^{-\frac{V_S}{2.5}}) + 2.5 ~{\rm log} (10^{-\frac{I_{cata}}{2.5}} - 10^{-\frac{I_S}{2.5}})\notag\\
            &= 2.57 \pm 0.54 \hspace{1cm} (V_S \equiv (V - I)_S + I_S)  \label{eq-VIb}
\end{align}
to $(V - I)_{\rm OGLE,map}$ in Equation (\ref{eq-cali2}) - (\ref{eq-cali3}) for $I_b$ calibration, Equation (\ref{eq-Ib}).

\subsection{Reddening and Extinction Correction}
\begin{figure}
\centering
\epsscale{0.7}
\rotatebox{-90}{
\plotone{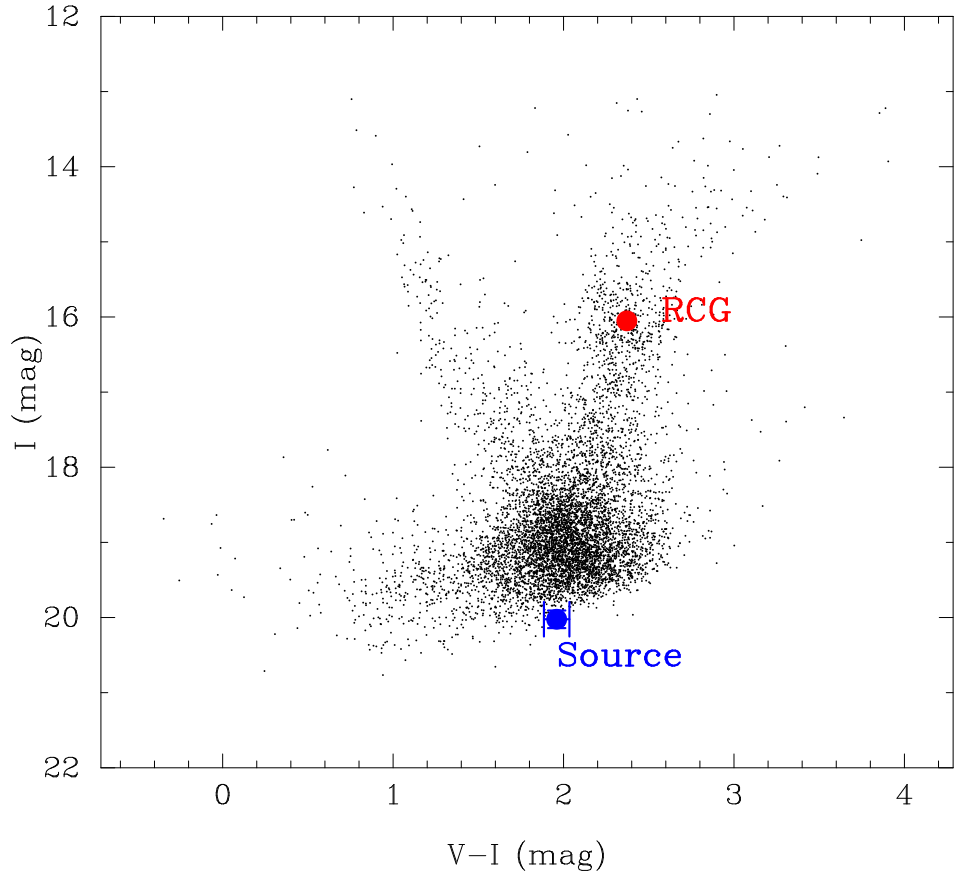}
}
\caption{The color-magnitude diagram of stars within $2^{\prime}$ of OGLE-2008-BLG-355 from the OGLE-III photometry map. The filled blue and red circles indicate the magnitude and color of the source and the central magnitude and color of the Red Clump Giants respectively.}
\label{fig-cmd}
\end{figure}

The source star magnitude and color need to be corrected for extinction and reddening due to the interstellar dust in the line of sight. We use red clump giants (RCG) as standard candles to estimate the extinction and reddening. The color-magnitude diagram (CMD) shown in Figure \ref{fig-cmd} is made from $V$- and $I$-band of the OGLE-III photometry map stars within 2$'$ around the source star. From CMD, we find the source star is likely to a G-type turn-off star in the Galactic bulge and the observed RCG centroid is
\begin{equation}
( V - I , I )_{RCG, obs} = ( 2.37, 16.07) \pm (0.01, 0.04).
\end{equation}
We adopt the intrinsic RCG color $( V - I )_{RCG, 0} = 1.06 \pm 0.06$ \citep{ben11} and the intrinsic RCG magnitude $I_{RCG, 0} = 14.44 \pm 0.04$ \citep{nat12} in this field,
\begin{equation}
( V - I , I )_{RCG, 0} = ( 1.06, 14.44 ) \pm ( 0.06, 0.04 ).
\end{equation}
We can derive the average reddening and extinction in this field by comparing the observed RCGs with intrinsic RCG color and magnitudes,
\begin{equation}
( E(V - I) , A_I ) = ( 1.31, 1.63) \pm ( 0.06, 0.06).
\end{equation}
The dereddened source color and magnitude are derived by applying these reddening and extinction values to the observed source color and magnitude given by Equation (\ref{eq-VIs}) (\ref{eq-Is}), 
\begin{equation}
( V - I , I )_{S, 0} = ( 0.66, 18.40 ) \pm ( 0.10, 0.13). \label{eq-VIs0} 
\end{equation}
The blending magnitudes of Equation (\ref{eq-Ib}) and (\ref{eq-VIb}) could be dereddened as well, 
\begin{align}
(V, I)_{b, 0} = (18.86, 17.60) \pm (0.60, 0.09). \label{eq-V,Ib0} \\
\end{align}

\subsection{The Angular Radii of the Source and Einstein Ring}
From $(V- I, I)_{S,0}$, we derive $(V- K, K)_{S,0} = (1.38, 17.65) \pm (0.18, 0.18)$ using a color-color relation \citep{bes88}.
Then, we apply a relation between $(V - K, K)_{S,0}$ and the stellar angular radius $\theta_*$
\citep{ker04} and estimate the source star angular radius $\theta_*$,
\begin{equation}
\theta_* = 0.62 \pm 0.07 \ \mu \rm{as}. 
\end{equation}
The angular Einstein radius $\theta_E$ and lens-source relative proper motion $\mu_{rel}$ are estimated, respectively, as
\begin{equation}
\theta_E = \frac{\theta_*}{\rho} = 0.28 \pm 0.03  \ \rm{mas},\label{eq-thetaE} 
\end{equation}

\begin{equation}
\mu_{rel} = \frac{\theta_E}{t_E} = 3.06 \pm 0.37 \ \rm{mas / yr}.\label{eq-murel} 
\end{equation}

\section{Lens System Masses and Distance}\label{sec-like}
\begin{figure}
\centering
\epsscale{0.7}
\rotatebox{-90}{
\plotone{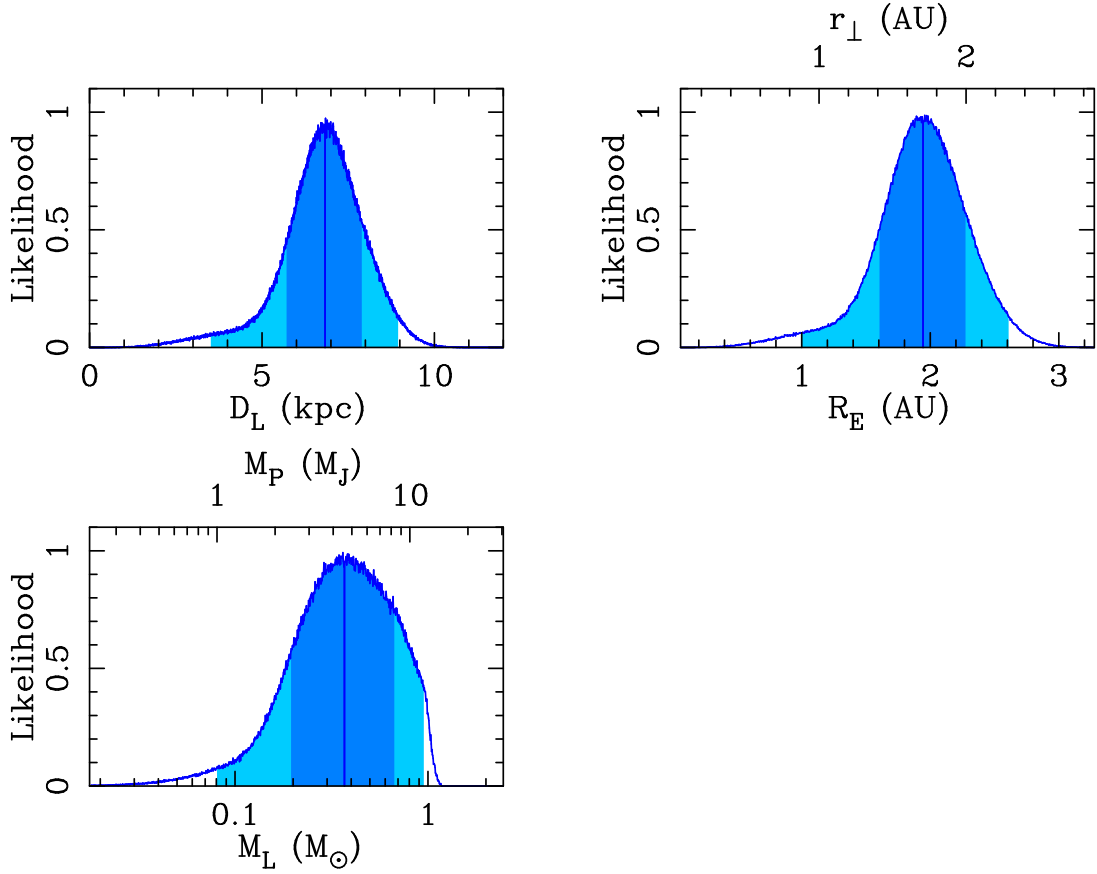}
}
\caption{Probability distributions from a Bayesian analysis constrained by $\theta_E, t_E$ and the upper limits of lens brightness, $I > I_{b,0}' = 19.04 \pm 0.47$ and $V > V_{b,0} = 18.86 \pm 0.60$, for distance to the lens, $D_L$, the mass of the lens system and the secondary, $M_{L}$, $M_{P} = q \times M_L$, the Einstein radius, $R_E$, and the projected separation, $r_\perp = R_E \times s$ of the lens system. The vertical solid lines indicate the median values. The dark and light shaded regions indicate the $1\sigma$ and $2\sigma$ limits.}
\label{fig-like}
\end{figure}

In this event, because the parallax effect is not detected, the lens system mass, $M_L$, and distance, $D_L$, is still degenerate according to the relation,
\begin{equation}
{\theta_E}^2 = \kappa M_{L} \left(\frac{1\rm{AU}}{D_L} + \frac{1\rm{AU}}{D_S}\right),\hspace{1cm} \kappa \equiv \frac{4G}{c^2\rm{AU}} \sim 8.14 \frac{\rm{mas}}{M_{\odot}} \label{eq-con}
\end{equation}
where the angular Einstein radius $\theta_E$, is given by Equation (\ref{eq-thetaE}) and $D_S$ is the distance to source, assumed to be located in the Galactic Center.
However, with our derived value of $\theta_E$, and our observed value of $t_E$, we can constrain the unknown event parameters by a Bayesian analysis using a model of Galactic kinematics \citep{alc95,bea06,gou06,ben08}.
We compute the likelihood by combining Equation (\ref{eq-con}) and the observed values of $\theta_E$ and $t_E$ with the Galactic model \citep{han03} assuming the distance to the Galactic Center is 8 kpc. Blending magnitudes can also be used in this calculation as the upper limit of lens brightness.
Because the brighter neighbor star seen in OGLE $I$-band image of Figure \ref{fig-nearby} cannot resolved, we consider that at least half of the baseline light comes from the neighbor. Therefore, we use
\begin{equation}
I_{b,0}' = 19.04 \pm 0.47 \label{eq-Ib,0'}
\end{equation}
obtained by subtracting the source brightness from the half of the baseline brightness as the upper limit of lens $I$ brightness for stronger constraint. Also, we use $V_{b,0}$ for the upper limit of lens $V$ brightness.
Figure \ref{fig-like} and Figure \ref{fig-like-color} show the likelihood distributions as the result of our analysis.  The primary star has the mass of $M_{\rm{h}} = 0.37^{+0.30}_{-0.17} M_{\odot}$, located at $D_L = 6.8_{-1.1}^{+1.1}$ kpc from the Earth and the planetary star is a gas giant with a mass of $M_{\rm{P}} = 4.6^{+3.7}_{-2.2} M_{\rm{J}}$ and a projected separation of $r_{\perp} = 1.70_{-0.30}^{+0.29} $AU.
The three-dimensional star-planet separation is statistically estimated to be $a = 2.0^{+1.0}_{-0.5}$AU by putting a planetary orbit at a random inclination and phase on the assumption that the shape of the orbit is a circle \citep{gou92}.
The probability distribution of $V$-, $I$-, $H$-, and $K$-band magnitudes of the lens primary star is shown in Figure \ref{fig-like-color}. 
From the estimated parameters, the primary lens star is likely to be a late type star in the Galactic Bulge, and it is consistent with the lens-source relative proper motion value given by Equation (\ref{eq-murel}), $\mu_{rel} = 3.06$ mas/yr, which favors that the lens is located in the bulge rather than the disk in which typically $\mu$ = 5 - 10 mas/yr.

\section{Discussion} \label{sec-disc}

This event was identified as a planetary event as a result of a systematic analysis 
of all past binary events observed by MOA prior to the 2013 bulge season. Previously, 
this event had been identified as a brown dwarf mass ratio binary lensing event 
using only OGLE data \citep{jar10}. Our systematic analysis finds that the best
fit model using only the OGLE data is also a planetary model, and this points to the
importance of a systematic analysis probing all of parameter space in order to
find the best fit model. 

Our Bayesian likelihood analysis, based on a standard Galactic model indicates that
the planet OGLE-2008-BLG-355Lb is a gas giant orbiting an M-dwarf or a late K-dwarf
at 1$\sigma$ confidence. But the host could also be a G-dwarf. Such a massive planet
with a mass ratio of $q = 0.0118 \pm 0.0006$, are predicted to be especially rare around low-mass
stars, like M-dwarfs \citep{lau04,ken08}. Thus, one might be tempted to conclude that
the existence of this planet is a challenge to the core accretion theory because an M-dwarf
host star is favored by the Bayesian analysis.

There is a flaw in this argument challenging the core accretion theory, however. Our Bayesian
analysis assumed that host stars of all masses were equally likely to host a planet with
the measured mass ratio, and so it could be that it is only this assumption that
challenges the core accretion theory. To really test a theory, we need to start with
a prior that is consistent with the theory, and then compare that prior to the data.
A statistical analysis with planet detection efficiencies would be required to 
do a serious test of the theory. However, there
has been no core accretion theory prediction of how the probability of hosting a 
planet of a given mass ratio should scale with the host star mass at the 
orbital separations probed by microlensing \citep{ida05}.

The solution to this problem is to determine the host star mass. For some events
\citep{gau08,ben08,mur11,kai13,pol13,tsa13,shv13}, this can be done with light curve measurements
of finite source effects and the microlensing parallax effect, but the OGLE-2008-BLG-355
light curve does not allow a measurement of the microlensing parallax effect. Fortunately,
lens star and planet masses can also be determined if the lens star is detected in
high angular resolution follow-up observations \citep{ben06,bennett07,don09a,benet10,kub12,bat14}.
In some cases, partial microlensing parallax information can be used to put constraints
on the lens system mass \citep{bat11}, or a partial microlensing parallax measurement can be combined with 
high angular resolution follow-up observations \citep{don09b} to yield a lens system
mass measurement. In the case of
the two-planet system OGLE-2006-BLG-109Lb,c, the microlensing parallax mass
measurement was confirmed by the host star detection in a high angular resolution
image \citep{benet10}. 

Two planetary events similar to OGLE-2008-BLG-355 are OGLE-2003-BLG-235
and MOA-2011-BLG-293. In both cases, a planet with a super-Jupiter mass ratio
($q \gg 0.001$) was found orbiting a star determined to be a likely M-dwarf by
a Bayesian analysis. In both cases, high angular resolution follow-up data was 
obtained after the event, and the follow-up data indicated that the lens stars
were near the upper end of the mass range allowed by the Bayesian analysis.
Neither host star turned out to be an M-dwarf. The host star OGLE-2003-BLG-235L
was determined to have a mass of $M_h = 0.63 {+0.07\atop -0.09}\msun$
\citep{ben06}, and the host star MOA-2011-BLG-293L was found to have 
a mass of $M_h = 0.86\pm 0.06 \msun$ \citep{bat14}.
This suggests that there may be some truth in the core accretion theory prediction
that massive gas giants are rare around M-dwarfs, particularly low-mass M-dwarfs.

\begin{figure}
\epsscale{0.7}
\rotatebox{-90}{
\plotone{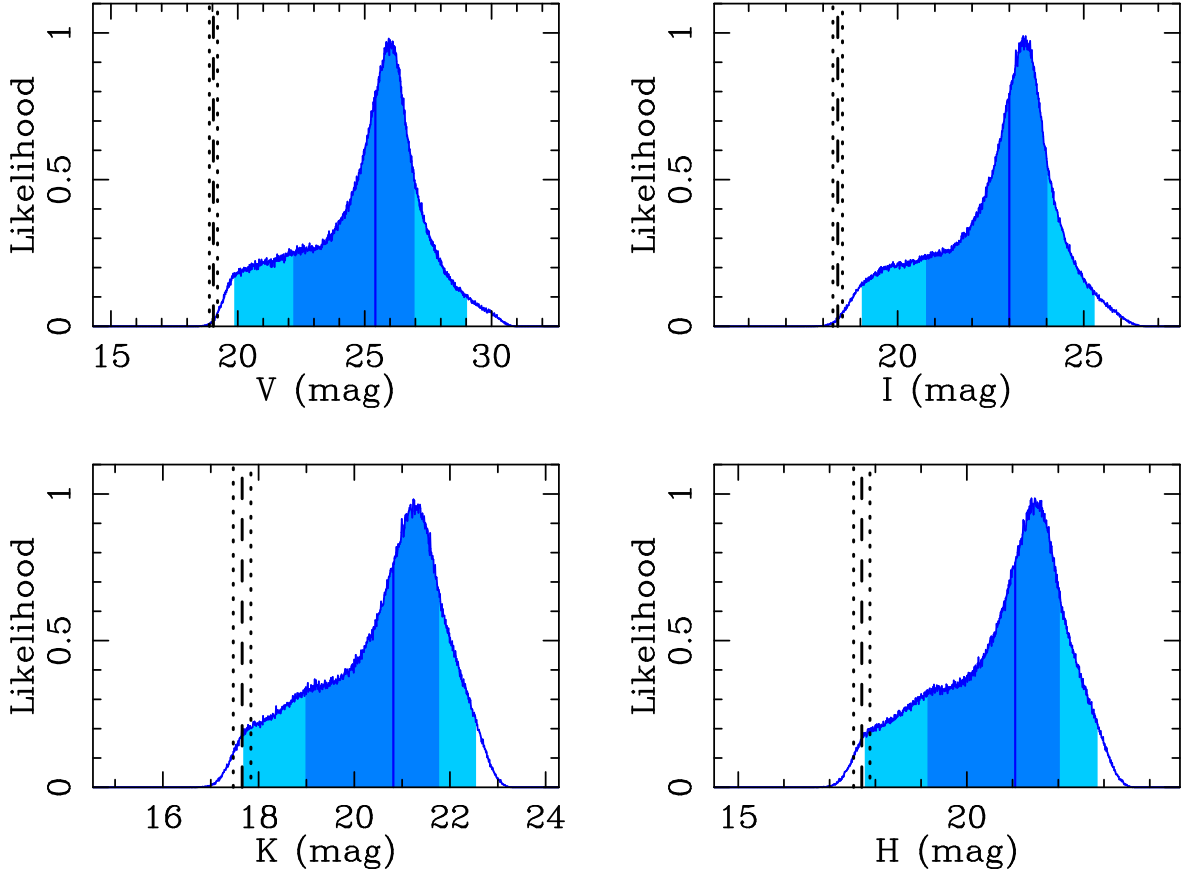}
}
\caption{Probability distributions from a Bayesian analysis constrained by $\theta_E, t_E$ and the upper limits of lens brightness, $I > I_{b,0}' = 19.04 \pm 0.47$ and $V > V_{b,0} = 18.86 \pm 0.60$, for $V$-, $I$-, $H$-, and $K$-band intrinsic magnitudes of the primary star of the lens system. The vertical solid lines indicate the median values. The dark and light shaded regions indicate the $1\sigma$ and $2\sigma$ limits. The black vertical dashed and dotted lines in each panel represent the source intrinsic magnitudes and $1\sigma$ error, respectively. The $K$- and $H$-band source magnitudes are derived using color-color relation in \citet{bes88}.}
\label{fig-like-color}
\end{figure}

The way to really test this core accretion theory prediction is to do a statistical
analysis using events that have mass determinations from microlensing parallax
measurements or high angular resolution follow-up observations that detect the
host star. Table \ref{tab-events} lists the microlensing events with host mass
determinations from either microlensing parallax or host star detection with
high angular resolution follow-up observations. Figure~\ref{fig-like-color}
shows that the host stars are likely to be within 3 magnitudes of the brightness
of the source star in the $H$ or $K$-bands, based on the Bayesian analysis of 
lens system properties. But, if the core accretion theory prediction is right,
then the lens star is likely to be on the bright side of the distributions in
Figure~\ref{fig-like-color}, and so the lens star would be easier to detect
than Figure~\ref{fig-like-color} implies.

This event is also one that was characterized using only MOA and OGLE data, which 
were the survey groups active in 2008. There are several other planetary events which 
are characterized without any data from follow up groups 
\citep{bon04,ben08,yee12,ben12,pol13,shv13,suz14} and 
these planets are all gas giants except MOA-2007-BLG-192Lb, 
which has relatively sparse coverage over caustic but fortuitously can be 
characterized \citep{ben08}. Note that follow-up observations with NACO adaptive 
optics system on the VLT was conducted for MOA-2007-BLG-192 and the 
refined physical parameters of the lens system \citep{kub12}
are consistent with the original results (see Table \ref{tab-events}).
MOA's normal observation cadence for the field containing OGLE-2008-BLG-355 
was every one observation per hour in 2008, but this event was characterized
thanks to increases in cadence by both OGLE and MOA in response to the
OGLE anomaly alert. At present, MOA has a 15 minute observing cadence in the
6 MOA fields ($13 {\rm deg}^2$) containing slightly more than half the microlensing events, while
the OGLE-IV survey 3 fields  ($4.2 {\rm deg}^2$) with a 20 minute cadence.
These observing cadences should enable us to detect perturbations due to smaller planets 
such as cold Neptunes or even Earth-mass planets \citep{gau12}. 
Therefore, it is expected that the type of planetary systems may in the future be found more using only survey data.

We acknowledge the following support: The MOA project was supported by a Grant-in-Aid for Scientific Research (JSPS19015005, JSPS19340058, JSPS20340052, JSPS20740104).
D.P.B.\ was supported by grants NASA-NNX12AF54G and NSF AST-1211875.
The OGLE project has received funding from the European Research Council under the European Community's Seventh Framework Programme (FP7/2007-2013)/ERC grant agreement no. 246678 to AU.





\clearpage



\clearpage


\clearpage

\def\@captype{table}
    \begin{center}
    \begin{tabular}{c|ccccccccccc}\hline\hline
Model & $\chi^2$ & $\chi^2/dof$ & $t_0$ & $t_E$ & $u_0$ & $q$ & $s$ & $\theta$ & $\rho$\\
      &        &          & (HJD') & (days) &  & ($10^{-2}$) & & (rad) & ($10^{-3}$)\\\hline
JA10 & 2257 & 1.566 & 4642.0 & 33.2 & 0.47 & 10.6 & 1.33  & 1.70 & 0.00\\\hline
Brown dwarf & 1539 & 1.071 & 4642.5 & 37.1 & 0.58 & 5.50 & 1.36  & 1.67 & 1.98\\\hline
Planetary   & 1444 & 1.005 & 4642.0 & 34.0 & 0.27 & 1.18 & 0.877 & 0.814 & 2.17\\
$\sigma$    &      &       &    0.2 &  2.2 & 0.03 & 0.06 & 0.010 & 0.022 & 0.15\\\hline\hline
    \end{tabular}
    \end{center}
 \tblcaption{The parameters of the best-fit planetary model and other models. Three models found correspond to the model in JA10. The model denoted "Brown dwarf" is the best model found with $q > 0.02$. The best model found in this work is the planetary model, with $\chi^2 = 1444$. The bottom row lists the 1$\sigma$ errors for the parameters of the planetary model.}
    \label{tab-models}
  \hfill

\clearpage

\tiny
\def\@captype{table}
    \begin{center}
    \begin{tabular}{c|cc|cc|c}\hline\hline
 & \multicolumn{2}{|c|}{Without follow-up data} &
\multicolumn{2}{|c|}{With follow-up data} & \\
Name & $M_{\rm h}$ & $M_{\rm P}$ & $M_{\rm h}$ & $M_{\rm P}$ & Paper\\
 & ($M_{\odot}$) &  & ($M_{\odot}$) &  & \\\hline
OGLE-2003-BLG-235L & $0.36^{+0.03}_{-0.28}$ & $1.5^{+0.1}_{-1.2} M_{\rm
J}$ & $0.63^{+0.07}_{-0.09}$ & $2.6^{+0.8}_{-0.6} M_{\rm J}$ &
\citet{bon04,ben06}\\
OGLE-2005-BLG-071L & 0.08-0.5 & 0.05-4 $M_{\rm J}$ & $0.46 \pm 0.04$ &
$3.8 \pm 0.4 M_{\rm J}$ & \citet{uda05,don09a}\\
OGLE-2006-BLG-109L & $0.50 \pm 0.05$ & $0.71 \pm 0.08 M_{\rm J}$ &
$0.51^{+0.05}_{-0.04}$ & $0.73 \pm 0.06 M_{\rm J}$ & \citet{gau08,benet10}\\
 &  & $0.27 \pm 0.03 M_{\rm J}$ &  & $0.27 \pm 0.02 M_{\rm J}$  & \\
OGLE-2009-BLG-151L & $0.018 \pm 0.001$ & $7.9 \pm 0.3 M_{\rm J}$ &  &  &
\citet{cho13}\\
OGLE-2011-BLG-0251L & $0.26 \pm 0.11$ & $0.53 \pm 0.21 M_{\rm J}$ &  &
& \citet{kai13}\\
OGLE-2011-BLG-0420L & $0.025 \pm 0.001$ & $9.9 \pm 0.5 M_{\rm J}$ &  &
& \citet{cho13}\\
OGLE-2012-BLG-0026L & $0.82 \pm 0.13$ & $0.11 \pm 0.02 M_{\rm J}$ &  &
& \citet{han13a}\\
 &  & $0.68 \pm 0.10 M_{\rm J}$ &  &  & \\
OGLE-2012-BLG-0358L & $0.022 \pm 0.002$ & $1.9 \pm 0.2 M_{\rm J}$ &  &
& \citet{han13b}\\
OGLE-2012-BLG-0406L & $0.44 \pm 0.07$ & $2.73 \pm 0.43 M_{\rm J}$ &  &
& \citet{tsa13}\\
MOA-2007-BLG-192L & $0.060^{+0.028}_{-0.021}$ & $3.3^{+4.9}_{-1.6}
M_{\oplus}$ & $0.084^{+0.015}_{-0.012}$ & $3.2^{+5.2}_{-1.8} M_{\oplus}$
& \citet{ben08,kub12}\\
MOA-2009-BLG-266L & $0.56 \pm 0.09$ & $10.4 \pm 1.7 M_{\oplus}$ &  &  &
\citet{mur11}\\
MOA-2010-BLG-328L & $0.11 \pm 0.01$ & $9.2 \pm 2.2 M_{\oplus}$ &  &  &
\citet{fur13}\\
MOA-2011-BLG-293L & $0.43^{+0.27}_{-0.17}$ & $2.4^{+1.5}_{-0.9} M_{\rm
J}$ & $0.86 \pm 0.06$ & $4.8 \pm 0.3 M_{\rm J}$ &
\citet{yee12,bat14}\\\hline\hline
    \end{tabular}
    \end{center}
 \tblcaption{The microlensing planetary systems whose masses are obtained without Bayesian analysis so far. Here, "follow-up" means a follow-up observation with high angular resolution by AO or a space telescope, not data from microlensing follow-up groups. Note that OGLE-2006-BLG-109L and OGLE-2012-BLG-0026L are multiple systems and the physical parameters of the former system were revealed without follow-up observations \citep{gau08}, and they are confirmed by Keck AO follow-up observation \citep{benet10}.}
    \label{tab-events}
  \hfill

\end{document}